\documentclass[aps,prc,twocolumn,superscriptaddress]{revtex4-1}

\usepackage{graphics}

\begin{document}


\title{Spectroscopy of $^{54}$Ti and the systematic behavior of low
  energy octupole states in Ca and Ti isotopes}



\author{L. A. Riley}
\affiliation{Department of Physics and Astronomy, Ursinus College,
  Collegeville, PA 19426, USA}

\author{M. L. Agiorgousis} 
\affiliation{Department of Physics and Astronomy, Ursinus College,
  Collegeville, PA 19426, USA}

\author{T.R. Baugher} 
\affiliation{National Superconducting Cyclotron Laboratory, Michigan
  State University, East Lansing, MI, 48824, USA}
\affiliation{Department of Physics and Astronomy, Michigan State
  University, East Lansing, MI, 48824, USA}

\author{D. Bazin} 
\affiliation{National Superconducting Cyclotron Laboratory, Michigan
  State University, East Lansing, MI, 48824, USA}

\author{R.L. Blanchard} 
\affiliation{Department of Physics and Astronomy, Ursinus College,
  Collegeville, PA 19426, USA}

\author{M. Bowry} 
\affiliation{National Superconducting Cyclotron Laboratory, Michigan
  State University, East Lansing, MI, 48824, USA}
\affiliation{Department of Physics and Astronomy, Michigan State
  University, East Lansing, MI, 48824, USA}

\author{P. D. Cottle} 
\affiliation{Department of Physics, Florida State University,
  Tallahassee, FL 32306, USA}

\author{F. G. DeVone} 
\affiliation{Department of Physics and Astronomy, Ursinus College,
  Collegeville, PA 19426, USA}

\author{A. Gade} 
\affiliation{National Superconducting Cyclotron Laboratory, Michigan
  State University, East Lansing, MI, 48824, USA}
\affiliation{Department of Physics and Astronomy, Michigan State
  University, East Lansing, MI, 48824, USA}

\author{M. T. Glowacki} 
\affiliation{Department of Physics and Astronomy, Ursinus College,
  Collegeville, PA 19426, USA}

\author{K. W. Kemper} 
\affiliation{Department of Physics, Florida State University,
  Tallahassee, FL 32306, USA}

\author{J.S. Kustina} 
\affiliation{Department of Physics and Astronomy, Ursinus College,
  Collegeville, PA 19426, USA}

\author{E. Lunderberg} 
\affiliation{National Superconducting Cyclotron Laboratory, Michigan
  State University, East Lansing, MI, 48824, USA}
\affiliation{Department of Physics and Astronomy, Michigan State
  University, East Lansing, MI, 48824, USA}

\author{D. M. McPherson} 
\affiliation{Department of Physics, Florida State University,
  Tallahassee, FL 32306, USA}

\author{S. Noji} 
\affiliation{National Superconducting Cyclotron Laboratory, Michigan
  State University, East Lansing, MI, 48824, USA}
\affiliation{Joint Institute for Nuclear Astrophysics - Center for the
  Evolution of the Elements, Michigan State University, East Lansing,
  MI 48824, USA}

\author{J. Piekarewicz} 
\affiliation{Department of Physics, Florida State University,
  Tallahassee, FL 32306, USA}

\author{F. Recchia} 
\altaffiliation{Dipartimento di Fisica e Astronomia “Galileo Galilei”,
  Universit‘a degli Studi di Padova, I-35131 Padova, Italy}
\affiliation{National Superconducting Cyclotron Laboratory, Michigan
  State University, East Lansing, MI, 48824, USA}

\author{B. V. Sadler} 
\affiliation{Department of Physics and Astronomy, Ursinus College,
  Collegeville, PA 19426, USA}

\author{M. Scott} 
\affiliation{National Superconducting Cyclotron Laboratory, Michigan
  State University, East Lansing, MI, 48824, USA}
\affiliation{Department of Physics and Astronomy, Michigan State
  University, East Lansing, MI, 48824, USA}

\author{D. Weisshaar} 
\affiliation{National Superconducting Cyclotron Laboratory, Michigan
  State University, East Lansing, MI, 48824, USA}

\author{R. G. T. Zegers} 
\affiliation{National Superconducting Cyclotron Laboratory, Michigan
  State University, East Lansing, MI, 48824, USA}
\affiliation{Department of Physics and Astronomy, Michigan State
  University, East Lansing, MI, 48824, USA}
\affiliation{Joint Institute for Nuclear Astrophysics - Center for the
  Evolution of the Elements, Michigan State University, East Lansing,
  MI 48824, USA}

\date{\today}

\begin{abstract}
Excited states of the $N=32$ nucleus $^{54}$Ti have been studied, via
both inverse-kinematics proton scattering and one-neutron knockout
from $^{55}$Ti by a liquid hydrogen target, using the GRETINA
$\gamma$-ray tracking array. Inelastic proton-scattering cross
sections and deformation lengths have been determined. A low-lying
octupole state has been tentatively identified in $^{54}$Ti for the
first time.  A comparison of $(p,p')$ results on low-energy octupole
states in the neutron-rich Ca and Ti isotopes with the results of
Random Phase Approximation calculations demonstrates that the observed
systematic behavior of these states is unexpected. 
\end{abstract}


\maketitle


\section{Introduction}

In atomic nuclei, collective low-energy octupole vibrations - like
other collective vibrations - are coherent sums of one particle-one
hole excitations.  Of course, to contribute to octupole vibrations the
particle-hole excitations must have $J^{\pi}=3^-$.  For low-energy
octupole states, that requires that a nucleon be promoted from a
common parity orbit to an opposite parity high-$j$ orbit
that has been depressed in energy due to the spin-orbit force.   

The occupation of the orbits involved in these particle-hole
excitations affects the systematic behavior of the octupole states in
a predictable way across most of the nuclear landscape.  For example,
the low-energy octupole state in the $N=Z=20$ isotope $^{40}$Ca is
composed primarily of excitations of both protons and neutrons from
$sd$ orbits to the $f_{7/2}$ orbit.  When neutrons are added to
$^{40}$Ca, they occupy the $f_{7/2}$ neutron orbit, blocking
$J^{\pi}=3^-$ excitations of neutrons from the $sd$ orbits.
Therefore, fewer particle-hole excitations are available to contribute
to the low-energy octupole state.  At $N=28$, the $f_{7/2}$ neutron
orbit is full and neutron contributions to the low-energy octupole
state are suppressed.  As a result, the energy of the $3_1^-$ state
increases from 3737 keV in $^{40}$Ca to 4507 keV in $^{48}$Ca. The
$E3$ excitation strength to the $3_1^-$ state as measured by electron
scattering decreases from $28(3)$ single particle units in $^{40}$Ca
to $9(2)$ single particle units in $^{48}$Ca.

By the same reasoning, adding neutrons to $^{48}$Ca should provide
additional particle-hole contributions to the low-energy octupole
state as excitations from the $p_{3/2}$ neutron orbit to the $g_{9/2}$
orbit become available.  We would expect that this would result in a
decrease in the energy of the low-energy octupole state and an
increase in $E3$ strength as the neutron number increases above
$^{48}$Ca.  However, a systematic experimental study of $3_1^-$ states
in the neutron-rich Ca isotopes by Riley \textit{et al.} \cite{Ril16}
found that is not the case.  Instead, the $(p,p')$ strength to the
$3_1^-$ states in $^{50,52}$Ca are equal to that in $^{48}$Ca, within
experimental uncertainties.

Here we report on a spectroscopic study of the exotic neutron-rich
isotope $^{54}$Ti via the reaction $(p,p')$ and knockout of a neutron
from $^{55}$Ti. Excited states of $^{54}$Ti have been
studies before via deep inelastic reactions of $^{48}$Ca on
$^{208}$Pb~\cite{Jan02} and $^{238}$U~\cite{For04} targets, beta
decay~\cite{Lid04, Cra10}, and Coulomb
excitation~\cite{Din05}. However, these reactions do not favorably
populate low-lying collective octupole states. The measurements of the
present work allowed us to determine the energy and strength
of the $3_1^-$ state of $^{54}$Ti and to examine the systematic
behavior of octupole states in the neutron-rich Ca and Ti isotopes.

To provide a theoretical context for these experimental results, we
also report the results of a Random Phase Approximation (RPA) study of
low-energy octupole states in the Ca and Ti isotopes.
  
\section{Experimental details}

The experiment was performed at the Coupled-Cyclotron Facility of the
National Superconducting Cyclotron Laboratory at Michigan State
University (NSCL). The secondary beam was produced by the
fragmentation of a 130~MeV/u $^{76}$Ge primary beam in a 376~mg/cm$^2$
$^9$Be production target and separated by the A1900 fragment
separator~\cite{A1900}. The momentum acceptance of the A1900 was
3\%. A 45~mg/cm$^2$ aluminum achromatic wedge was used to further
purify the secondary beam by $Z$.

\begin{figure}
  \scalebox{0.5}{
    \includegraphics{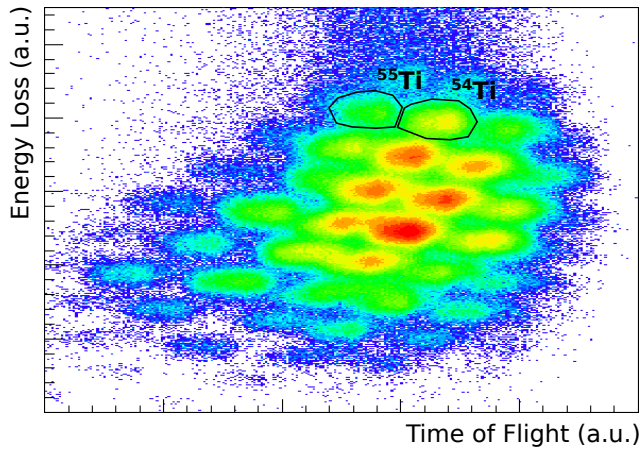}
  }
  \caption{\label{ti54_pid} (Color online) Energy loss in a Si PIN
    diode vs. time of flight from the A1900 extended focal plane to
    the S800 object scintillator of the incoming beam used for particle
    identification.}
\end{figure}

Secondary beam particles were identified upstream of the reaction
target by energy loss in a silicon surface barrier detector and by
time of flight from the A1900 extended focal plane to the S800 object
scintillator. The incoming particle identification spectrum is shown 
in Figure~\ref{ti54_pid}.
The beam then passed through the NSCL/Ursinus College
Liquid Hydrogen Target, based on the design of Ryuto et
al.~\cite{Ryu05}. The target was installed at the target position of
the S800 magnetic spectrograph~\cite{S800}. Beam-like reaction
products were identified downstream of the target by energy loss in
the S800 ion chamber and time of flight from the S800 object
scintillator to a scintillator in the focal plane of the S800.
The secondary beam contained components spanning the range
$14 \leq Z \leq 23$, including $^{54,55}$Ti. A total of
$1.59 \times 10^7$ $^{54}$Ti and $8.71 \times 10^6$ $^{55}$Ti
particles passed through the target during the measurement at rates of
39~particles/s and 21~particles/s, respectively. 

The liquid hydrogen was contained by a cylindrical aluminum target
cell with 125~$\mu$m Kapton entrance and exit windows mounted on a
cryocooler. The nominal target thickness was 30~mm.  The target cell
and cryocooler were surrounded by a 1~mm thick aluminum radiation
shield with entrance and exit windows covered by 5~$\mu$m aluminized
Mylar foil. The temperature and pressure of the target cell was
16.00(25)~K and 868(10)~Torr throughout the experiment. The variation
in the temperature and pressure of the target cell correspond to a
0.3~\% uncertainty in target density.

The Kapton windows of the target cell bulged significantly due to the
pressure difference between the cell and the evacuated beam line. The
effective thickness of the target was determined to be 258~mg/cm$^2$
by fitting \textsc{geant4} simulations of the beam passing through the target
to the measured kinetic energy distribution of the outgoing reaction
products as described in Ref.~\cite{Ril14}. The mid-target beam
energy was 91.5~MeV/nucleon. 

The GRETINA $\gamma$-ray tracking array~\cite{GRETINA, GRETINA2},
consisting of 28 36-fold segmented high purity germanium crystals
packaged in seven clusters of four crystals each, was installed at the
pivot point of the S800. In order to accomodate the liquid hydrogen
target, the array was mounted on the north half of the mounting shell
with two modules centered at 58$^\circ$, four at 90$^\circ$, and one
at 122$^\circ$ with respect to the beam axis.

\section{Experimental results}

\begin{figure}
  \scalebox{0.6}{
    \includegraphics{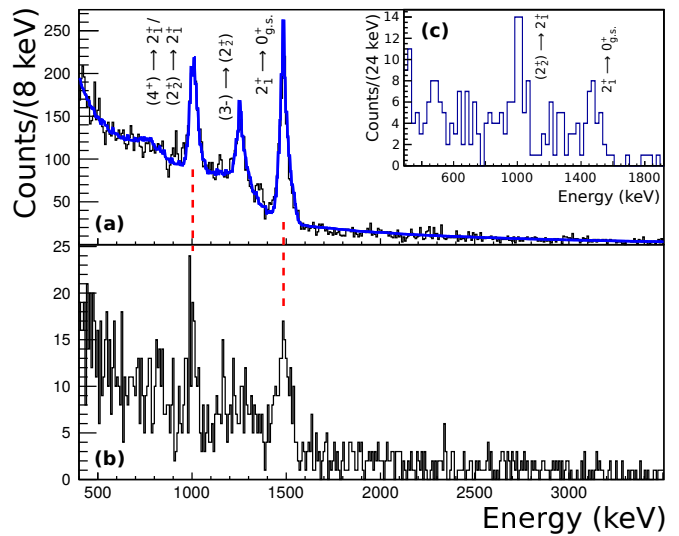}
  }
  \caption{\label{ti54_spectra} (Color online) Projectile-frame
    $\gamma$-ray spectra of $^{54}$Ti measured via (a)
    inverse-kinematics proton scattering and (b) one-neutron knockout
    from $^{55}$Ti. The solid curve is the fit of \textsc{geant4}
    simulations to the spectrum obtained from $(p,p')$. (c) Spectrum
    of $\gamma$ rays obtained from $(p,p')$ gated on the 1264-keV
    $\gamma$ ray.}
\end{figure}

\begin{figure}
  \scalebox{0.4}{
    \includegraphics{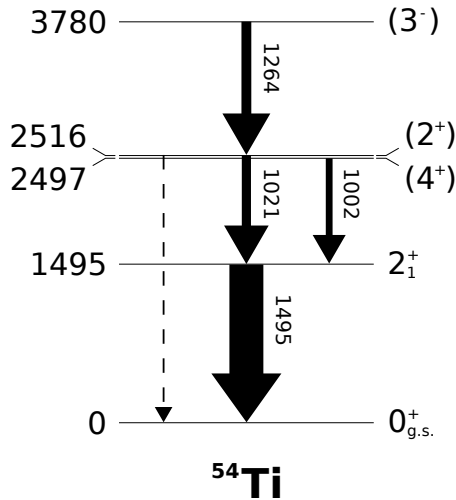}
  }
  \caption{\label{ti54_level}  Partial level scheme of $^{54}$Ti
  showing levels populated in the present work. Arrow widths are
  proportional to the $\gamma$-ray intensities measured in
  the $(p,p')$ reaction.}
\end{figure}

The Doppler-corrected spectrum of $\gamma$ rays measured in
coincidence with incoming and outgoing $^{54}$Ti particles,
corresponding to the $(p,p')$ reaction, is shown in panel (a) of
Figure~\ref{ti54_spectra}. Panel (b) of the figure shows the
$\gamma$-ray spectrum measured in coincidence with incoming $^{55}$Ti 
and outgoing $^{54}$Ti particles, corresponding to the one-neutron
knockout reaction. The average projectile speed used in Doppler
reconstruction was $v/c = 0.41$. The $\gamma$ rays at 1495, 1002, and
1021 keV correspond to transitions observed in the $\beta$-decay study
of Crawford et al~\cite{Cra10}. An additional $\gamma$-ray at 1264~keV
is observed for the first time. The $\gamma$-ray spectrum from $(p,p')$
measured in coincidence with the 1264-keV $\gamma$ ray is shown in
panel (c) of Figure~\ref{ti54_spectra}. On this basis, we identify the
1264-keV transition as part of a cascade involving the 1495-keV
$2^+_1 \rightarrow 0^+_\mathrm{g.s.}$ transition and one of the
transitions in the unresolved 1002/1021-keV doublet.

\begin{table*}
\caption{\label{tab:gammas} Level energies, spins and parities,
  $\gamma$-ray energies, and branching ratios (BR) from
  Ref.~\cite{Cra10}, $\gamma$-ray energies, intensities relative to
  that of the $2^+_1 \rightarrow 0^+_\mathrm{g.s.}$ transition, and
  branching ratios (BR), and cross sections from the present work.}
\begin{ruledtabular}
\begin{tabular}{rcclcccc}
& & \multicolumn{2}{c}{Ref.~\cite{Cra10}} 
                        & \multicolumn{4}{c}{Present work}\\
\cline{3-4}\cline{5-8}
$E_\mathrm{level}$ [keV] & $J^\pi$ [$\hbar$] & $E_\gamma$ [keV] &BR [\%]&
  $E_\gamma$ [keV] & $I_\gamma$ [\%] & BR [\%] & $\sigma$ [mb] \\\hline\hline
1495.0(3) & $2^+_1$   & 1495.0(3) &        & 1495(2)    & 100   &       & 8.7(8) \\
2497.4(4) & $(4^+)$   & 1002.4(3) &        & 1000(3)    & 20(3) &       & 3.3(4) \\
2517.1(3) & $(2^+_2)$ & 1020.8(4) & 64(14) & 1022(3)    & 27(3) & $>$84 & $<$0.7 \\
          &           & 2517.5(3) & 36(7)  & 2517    & $<$6  & $<$16 &        \\
3780(3)   & $(3^-)$   &           &        & 1264(3) & 29(3) &       & 4.9(4) \\
\end{tabular}
\end{ruledtabular}
\end{table*}

The solid curve overlayed on the $\gamma$-ray spectrum collected via
$(p,p')$ in panel (a) of Figure~\ref{ti54_spectra} is a fit of \textsc{geant4}
simulations of the response of GRETINA to the observed $\gamma$ rays
and prompt and non-prompt background using the method described in 
Ref.~\cite{Ril14}. The $\gamma$-ray energies and intensities
extracted from the fits are listed in Table~\ref{tab:gammas} along
with the branching ratios. Also included are level energies,
$\gamma$-ray energies and branching ratios from Ref.~\cite{Cra10}. 

The 2517-keV $(2^+_2) \rightarrow 0^+_{g.s.}$ transition observed in
the beta-decay study of Crawford et al.~\cite{Cra10} was included in
the fit. The best-fit yield of this $\gamma$ ray is consistent with
zero. The upper limit on its intensity in Table~\ref{tab:gammas} is
based on the statistsical uncertainty from the fit. A fit assuming the
branching ratio to the 2517-keV transition observed in
Ref.~\cite{Cra10} is not consistent with our spectrum.

A partial level scheme for $^{54}$Ti including states populated in the
present work is shown in Figure~\ref{ti54_level}.  The level scheme,
with the exception of the new state at 3780~keV, corresponds to that
established in Ref.~\cite{Cra10}. We place the 1264-keV transition
feeding the $(2^+_2)$ state at 2497~keV on the basis that the measured
relative intensity of the 1264-keV $\gamma$ ray is significantly
greater than that of the 1002-keV $\gamma$ ray.  Using this level
scheme, we applied feeding corrections to the measured $\gamma$-ray
yields to determine the cross sections for populating excited states
via proton scattering that are listed in Table~\ref{tab:gammas}.

We make a tentative assignent of $J^\pi = 3^-$ to the state at
3780~keV because it has a large $(p,p')$ cross section, it decays to
the $(2_2^+)$ state and it is in the broad energy range associated
with low-energy octupole states in this mass region~\cite{Ki02, Ril16}. 

The absence of the 1264-keV $\gamma$ ray from the neutron knockout
spectrum is consistent with the $3_1^-$ assignment for the 3780-keV
state for reasons that will be explained below.

We used the coupled-channels code \textsc{ECIS95}~\cite{Ray95} and the
global optical potential of Ref.~\cite{Kon03} to determine deformation
lengths from our measured cross sections for inelastic scattering to
the $2^+_1$, $3^-$ and $4_1^+$ states of $\delta_2 = 0.93(4)$~fm,
$\delta_3 = 0.76(3)$~fm and $\delta_4=0.67(4)$~fm.  

\section{Random Phase Approximation Calculations}

The distribution of isoscalar $E3$ strength is computed in a
relativistic random phase approximation (RPA) using three effective
interactions that have been calibrated to the properties of finite
nuclei: NL3 \cite{La97,La99}, FSUGold \cite{To05}, and FSUGarnet
\cite{Ch15}.  The development of FSUGold used NL3 as a starting point, 
and FSUGarnet is a refinement of FSUGold that
has been fitted to a few giant monopole energies and to
well-established properties of neutron stars \cite{Ch14}.  Moreover,
FSUGarnet predicts that the isotopic chain in oxygen can be made to
terminate at $^{24}$O, as has been observed experimentally
\cite{Th04}.  The parameters for the three relativistic models are
tabulated in \cite{Pi17}.

The implementation of the RPA formalism adopted here is rooted in a
non-spectral approach that allows for an exact treatment of the
continuum without any reliance on discretization.  This is
particularly critical in the case of weakly-bound nuclei with
single-particle orbits near the continuum. The discretization of the
continuum is neither required nor admitted.   The relativistic RPA
formalism as implemented here has been reviewed extensively in earlier
publications; see for example  Refs. \cite{Pi14,Pi15} and references
contained therein.  In the context of the isoscalar monopole response
of the neutron-rich calcium isotopes, the details of the formalism are
described in \cite{Pi17}. 

\begin{figure}
  \scalebox{1.0}{
    \scalebox{0.55}{
      \includegraphics{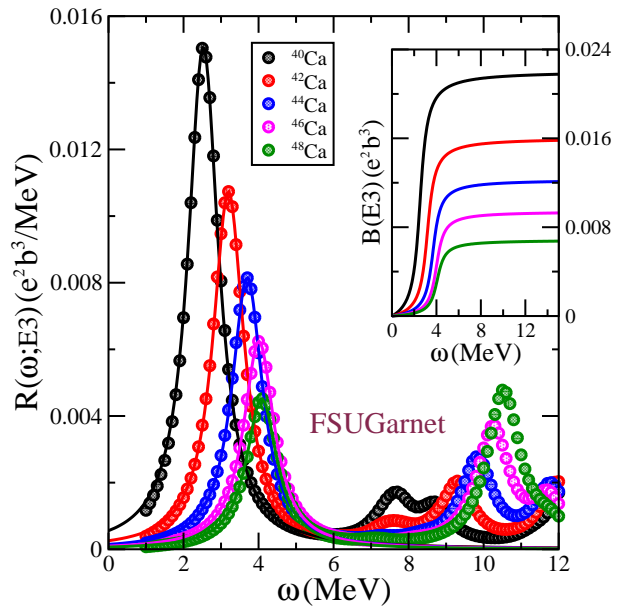}
    }
  }
  \caption{\label{ca4048_rpa} (Color online) $E3$ strength distributions for
    $^{40,42,44,46,48}$Ca calculated with the RPA.}
\end{figure}

The RPA calculations were performed for $^{40,42,44,46,48,50,52}$Ca
and $^{50,52,54}$Ti.  Figure \ref{ca4048_rpa} shows the $E3$ strength
distributions calculated for $^{40,42,44,46,48}$Ca using FSUGarnet, which
is the most refined of the three interactions used here.  The first feature
to notice is that in each of these isotopes there is a concentration
of strength below 6 MeV.  This is the Low-Energy Octupole State (LEOS)
that is sometimes - but not always - concentrated in the $3_1^-$
state. 

Above 6 MeV, a second concentration of $E3$ strength can be seen - the
Low-Energy Octupole Resonance (LEOR) - which occurs throughout the
Periodic Table at energies near (31 MeV)/$A^{1/3}$ \cite{Ki82}.

Both the LEOS and LEOR are composed of $1 \hbar \omega$ one
particle-one hole excitations.  However, the LEOS is composed of
excitations to the high-spin ``intruder" orbit that is pushed down by
the spin-orbit force from the major shell $1 \hbar \omega$ above.  The
LEOR includes excitations to the rest of the major shell $1 \hbar
\omega$ above. 

For $N<28$ Ca isotopes, the LEOS is composed mostly of excitations of
both protons and neutrons from the $sd$ shell to the $f_{7/2}$ orbit.
The LEOR is mostly composed of excitations from  the $sd$ orbits to
the $p_{3/2}$, $p_{1/2}$ and $f_{5/2}$ orbits. 

In Figure \ref{ca4048_rpa}, the RPA calculation clearly shows the
behavior we expect in the simplest picture of octupole vibrations.
The maximum number of excitations from the $sd$ shell to the $f_{7/2}$
orbits is available in $N=Z=20$ $^{40}$Ca.  Therefore, the strength of
the LEOS is at a maximum and the energy of the LEOS is at a minimum in
that isotope.  As neutrons are added to $^{40}$Ca, the increasing
occupation of the $f_{7/2}$ neutron orbit blocks neutron contributions
to the LEOS.  As a result, the strength of the LEOS decreases and its
energy increases until $^{48}$Ca is reached and the LEOS energy
reaches a maximum and the $E3$ strength reaches a minimum.

\section{Systematic Behavior of Low-Energy Octupole States in
  C\lowercase{a} and T\lowercase{i} isotopes}   

\begin{figure}
  \scalebox{1.0}{
    \scalebox{0.55}{
      \includegraphics{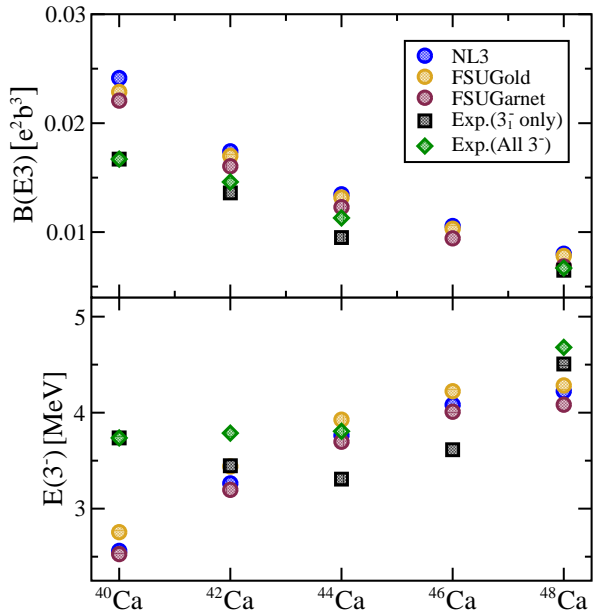}
    }
  }
  \caption{\label{ca_syst}  (Color online) Comparison of the systematic behavior of
    RPA calculations of LEOS energy and $E3$ strength for
    $^{40,42,44,46,48}$Ca with data.  The black squares show the
    energies of the $3_1^-$ states and the $B(E3;\uparrow)$ values for
    the $3_1^-$ states from electron scattering.  The green diamonds
    show the total $E3$ strength for all $3^-$ states observed in
    electron scattering below 6 MeV, and the centroids of this $E3$
    strength.} 
\end{figure}

Before we examine the results of the calculations for the Ti isotopes,
including $^{54}$Ti which is the focus of the present work, we compare
the results of the RPA calculations for Ca isotopes to the data.
Figure \ref{ca_syst} compares the RPA results using the three models
to the energies and $E3$ strengths of the LEOS in
$^{40,42,44,46,48}$Ca. The energies of states shown in Figure
\ref{ca_syst} are taken from compilations
\cite{Ch17,Ch16,Ch11,Wu00,Bu06}.  The $E3$ strengths are from electron
scattering (also taken from the compilations), which provides a
uniform data set for these isotopes.  The points labeled ``All $3^-$"
are explained in the next paragraph. 

If we only look at the $3_1^-$ states in Figure \ref{ca_syst}, we see
an energy trend that we do not expect for the LEOS.  As neutrons are
added to $^{40}$Ca, the energy of the $3_1^-$ state decreases.
However, the electron scattering measurements revealed other $3^-$
states in the LEOS energy range (below 6 MeV) in $^{42,44,48}$Ca. That
is, it appears that the LEOS has fragmented in these three isotopes
(there are no electron scattering results for $^{46}$Ca).  If instead
of only considering the $3_1^-$ states we say that the LEOS energies 
are the centroids of the strength as seen in electron scattering, then
the decrease in energy does not occur, as indicated with ``All $3^-$"  in
Figure \ref{ca_syst}.

Figure \ref{ca_syst} also shows the LEOS $E3$ strength results both
without accounting for LEOS fragmentation (that is, only the $3_1^-$
states) and including the LEOS fragments observed in electron
scattering.  For both LEOS energies and strengths, the calculations
qualitatively reproduce the trends in the data for
$^{40,42,44,46,48}$Ca.

\begin{figure}
  \scalebox{1.0}{
    \scalebox{0.55}{
      \includegraphics{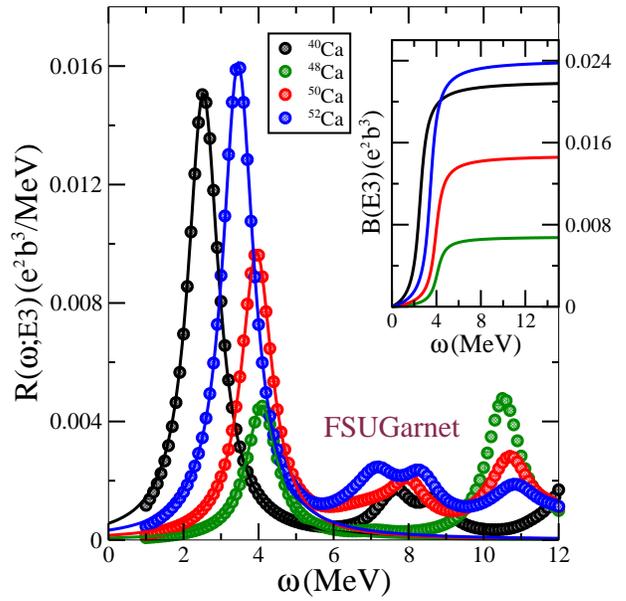}
    }
  }
  \caption{\label{ca4852_rpa} (Color online) $E3$ strength distributions for
    $^{40,48,50,52}$Ca calculated with the RPA.}
\end{figure}

The RPA strength distributions for $^{48,50,52}$Ca are shown in Figure
\ref{ca4852_rpa} and compared to the RPA result for $^{40}$Ca.  The
RPA calculation predicts that when neutrons are added to $^{48}$Ca the
$E3$ strength of the LEOS increases quickly and the energy declines.
The calculated $E3$ strength increases by more than a factor of three
from $^{48}$Ca to $^{52}$Ca, reaching a value comparable to that in
$^{40}$Ca.

The origin of the increase in $E3$ strength and decrease in LEOS
energy as neutrons are added to $^{48}$Ca in the RPA calculation is
straightforward to understand.  The neutrons added to $^{48}$Ca mostly
occupy the $p_{3/2}$ orbit, adding one neutron-one neutron hole
excitations to the $g_{9/2}$ orbit for the LEOS wavefunction.  We
note, however, that the $g_{9/2}$ orbit is predicted to be in the
continuum, so the non-spectral approach adopted here offers a distinct
advantage over the spectral approach.

\begin{figure}
  \scalebox{1.0}{
    \scalebox{0.55}{
      \includegraphics{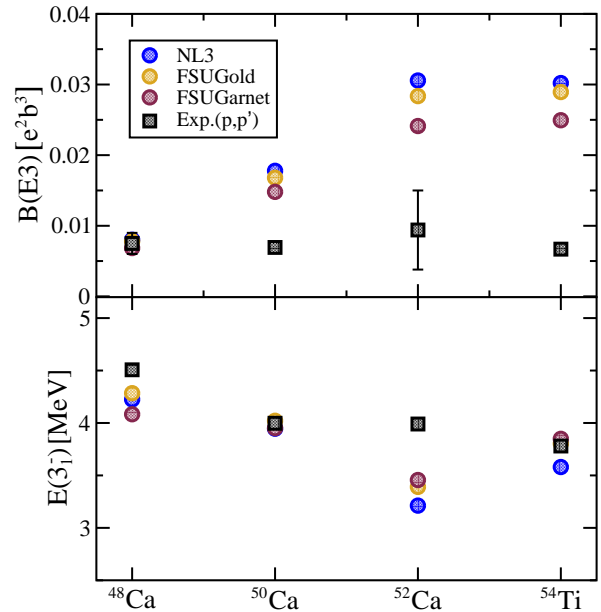}
    }
  }
  \caption{\label{ti_syst} (Color online) Comparison of the systematic behavior of
    RPA calculations of LEOS energy and $E3$ strength for
    $^{48,50,52}$Ca and $^{54}$Ti with data.  The black squares show
    the energies of the $3_1^-$ states and the $B(E3;\uparrow)$ values
    for the $3_1^-$ states from proton scattering in inverse
    kinematics (Ref. \cite{Ril16} and the present work).}
\end{figure}

Figure \ref{ti_syst} compares the RPA results with data on octupole
states in $^{48,50,52}$Ca from Ref. \cite{Ril16}, in which these
isotopes were studied via the $(p,p')$ reaction in inverse kinematics
- providing a uniform data set for all three nuclei.  The data show
that the energies of the $3_1^-$ states drop from 4.5 MeV in $^{48}$Ca
to 4.0 MeV in $^{50,52}$Ca. The $E3$ strength for exciting the $3_1^-$
states stays constant within experimental uncertainty.  The
$B(E3;\uparrow)$ values are calculated from the $\delta_3$ values in
Ref. \cite{Ril16} using the prescription given in Ref. \cite{Ki02} and
the real radius parameters from Ref. \cite{Kon03}.  The $E3$ strength
is constant, in contrast with the RPA prediction that rises sharply.
So the experimental results for $^{50,52}$Ca can be considered
unexpected. 

However, it is important to note that the radioactive beam experiment
from which the results in Figure \ref{ti_syst} were extracted only
yielded the observation of a single $3^-$ state in each nucleus.  It
is possible that the LEOS is fragmented in $^{50,52}$Ca and that this
fragmentation was not observed in the experiment.

\begin{figure}
  \scalebox{1.0}{
    \scalebox{0.55}{
      \includegraphics{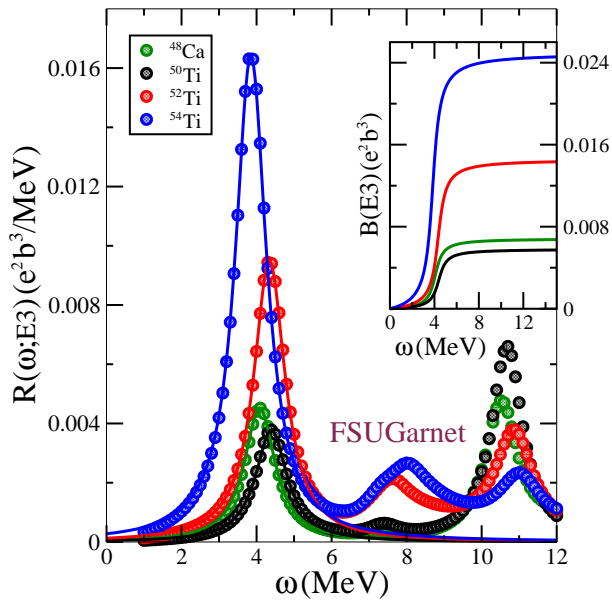}
    }
  }
  \caption{\label{ti_rpa} (Color online) $E3$ strength distributions for
    $^{48}$Ca and $^{50,52,54}$Ti calculated with the RPA.}
\end{figure}

Figure \ref{ti_rpa} shows the results of RPA calculations for
$^{48}$Ca and $^{50,52,54}$Ti.  The calculations predict that the
$N=28$ isotope $^{50}$Ti has a somewhat smaller $E3$ strength in the
LEOS than $^{48}$Ca because the two additional protons occupy the
$f_{7/2}$ orbit and block some of the one proton-one hole excitations
from the $sd$ shell that contribute to the LEOS in $^{48}$Ca.
However, a much larger change to the $E3$ strength occurs when
neutrons are added to $^{50}$Ti:  the calculation predicts that the
$E3$ strength in $^{54}$Ti is four times higher than in $^{50}$Ti.  As
in the $N>28$ Ca isotopes, that rapid increase in the calculated $E3$
strength is driven by the occupation of the $p_{3/2}$ neutron orbit
which makes available one neutron-one hole excitations to the unbound
$g_{9/2}$ orbit.

However, the present $(p,p')$ results show that the $E3$ strength to
the $3_1^-$ state in $^{54}$Ti is approximately equal to the
corresponding strength in $^{48}$Ca, as shown in Figure \ref{ti_syst}.
Once again, the calculations can be understood in a straightforward
way and we can conclude that the systematic behavior revealed by the
state is unexpected unless an unobserved strong fragmentation of the strength occurs.

This explanation of the predicted role of the $p_{3/2}$-$g_{9/2}$
neutron-neutron hole contribution to the $3_1^-$ state in $^{54}$Ti
provides an explanation for the absence of the $3_1^-$ state from the
$^{55}$Ti knockout spectrum.  The neutron knockout reaction would only
populate the $3_1^-$ state if the ground state of $^{55}$Ti has a
neutron in the $g_{9/2}$ orbit.  In the ground state of $^{55}$Ti, the
neutrons occupy the $fp$ orbits, so the knockout reaction cannot
populate the $3_1^-$ state.

The fragmentation of the LEOS in $^{42,44}$Ca may suggest an
explanation for the anomalous behavior of $3_1^-$ states in
$^{50,52}$Ca and $^{54}$Ti.  The challenging nature of the exotic beam
experiments reported here make it possible that fragments of the LEOS
- particularly those lying higher in energy than the $3_1^-$ states
reported here and in Ref. \cite{Ril16} - were not detected.  If the
LEOS is fragmented in $^{50,52}$Ca and $^{54}$Ti, the total $E3$
strength to the LEOS might be considerably larger than we have
observed in the $3_1^-$ states of these nuclei.

It is worth noting that $^{42,44}$Ca are not the only nuclei in this
mass region in which the LEOS is significantly fragmented.  Higashi
\textit{et al.} \cite{Hi89} used 65 MeV proton scattering to observe
the fragmentation of the LEOS in stable $^{46,48}$Ti.  We can conclude
that the fragmentation of the LEOS is the most likely explanation for
the large discrepancies between the experimental $(p,p')$ results for
$3_1^-$ states in $^{50,52}$Ca and $^{54}$Ti and the RPA calculations
presented here.
      
\section{Summary}

The spectroscopy of $^{54}$Ti has been performed with two reactions in
the same target - inelastic proton scattering in inverse kinematics
and one-neutron knockout from $^{55}$Ti.  A state at 3780 keV has
been tentatively identified as the $3_1^-$ state.  Deformation lengths
have been extracted from the $(p,p')$ data for the $2_1^+$, $4_1^+$
and $3_1^-$ states.

RPA calculations have been performed for the LEOS in
$^{40,42,44,46,48,50,52}$Ca and $^{50,52,54}$Ti.  The observed $E3$
strength in the $3_1^-$ states of $^{50,52}$Ca and $^{54}$Ti is
considerably smaller than the predicted values.  This may be because
the LEOS is fragmented in these isotopes.  A more sensitive experiment
to detect other $3^-$ states in these isotopes should be performed to
determine whether fragmentation causes the discrepancy between the
data and the calcuation or whether the strength of the LEOS is
quenched.

\begin{acknowledgments}
This work was supported by the National Science Foundation under Grant
Nos.  PHY-1617250, PHY-1303480, PHY-1064819, PHY-1565546, and
1401574.  One of us (JP) is supported by the U.S. Department of
Energy Office of Science, Office of Nuclear Physics under Award Number
DE-FD05-92ER40750. GRETINA was funded by the US DOE - Office of
Science. Operation of the array at NSCL was supported by NSF under
Cooperative Agreement PHY-1102511(NSCL) and DOE under grant
DE-AC02-05CH11231(LBNL). We also thank T.J. Carroll for the use of the
Ursinus College Parallel Computing Cluster, supported by NSF grant
no. PHY-1205895.
\end{acknowledgments}


%

\end{document}